\documentclass[11pt,twoside]{article}
\usepackage{asp2010}
\usepackage{epsf}
\usepackage{lscape}
\usepackage{natbib}
\usepackage{graphicx}

\begin{document}
\bibliographystyle{asp2010}

\title{The nonlinear evolution of large scale structures in Growing Neutrino cosmologies}

\author{Marco Baldi
  \affil{Excellence Cluster Universe, Boltzmannstr.~2, D-85748 Garching, Germany
\\University Observatory, Ludwig-Maximillians University Munich, Scheinerstr. 1, D-81679 Munich, Germany}
  }

\begin{abstract}
We present the results of the first N-body simulations of the Growing Neutrino scenario, as 
recently discussed in \citet{Baldi_etal_2011a}. Our results have shown for the first time how
neutrino lumps forming in the context of Growing Neutrino cosmologies are expected to pulsate
as a consequence of the rapid oscillations of the dark energy scalar field. We have also computed for the
first time a realistic statistical distribution of neutrino halos and determined their impact on the underlying Cold Dark Matter
structures.
\end{abstract}


\section{Introduction: the Growing Neutrino scenario}

The Growing Neutrino scenario has been recently proposed by \citet{Amendola_Baldi_Wetterich_2008} as a possible solution to the so called
``cosmic coincidence" problem, i.e. the fact that the two main components that constitute our Universe, Dark Energy (DE) and Cold Dark Matter (CDM),
happen to have comparable densities only around the present cosmological epoch.
The Growing Neutrino scenario is based on a direct interaction between a DE scalar field $\phi $, which is considered to be responsible for the late-time acceleration of the expansion of the universe instead of the standard cosmological constant,
and massive neutrinos. This interaction determines an exchange of energy between the two components such that the individual energy-momentum tensors of DE and
massive neutrinos are not separately conserved, while only their sum obeys the standard conservation equation provided by General Covariance.
This exchange of energy is encoded in the dependence of the average neutrino mass on the global evolution of the DE scalar field $\phi $, according to the equation:
\begin{equation}
\label{eq:nu_mass} 
m_{\nu} \equiv m_{\nu}(t_{0}) e^{-{{\beta}} \phi} \,,
\end{equation}
where $m_{\nu}(t_{0})$ is a constant that represents the value of the neutrino mass at the present time and the parameter $\beta $ sets the strength of the coupling. 
The dynamics of the DE scalar field $\phi $ will therefore determine a time dependence of the neutrino mass, and for a given set of cosmological
parameters the background evolution of the system can be computed.
In particular, a specific feature of viable Growing Neutrino models (characterized by a large and negative value of the coupling $\beta $, which is taken to be $\beta =-52$
for the realization considered in the present work) is the sudden growth of the neutrino mass at the time when neutrinos become non-relativistic particles. 
In fact, while for relativistic particles the effect of the coupling is suppressed by pressure terms in the neutrino continuity equation,
as soon as the neutrino equation of state parameter $w_{\nu }\equiv p_{\nu }/\rho _{\nu }$ drops from its relativistic value of $1/3$ to zero the coupling becomes active 
and introduces a steep minimum in the total effective scalar potential where the scalar field $\phi $ rapidly oscillates and eventually stops.

The trapping of the DE scalar field in a steep local minimum of its effective potential determines a sudden drop of the DE equation of state $w_{\phi }$ towards the
value of $-1$, and the DE scalar field therefore starts to behave as a cosmological constant. The transition from relativistic to non-relativistic neutrinos is therefore
the trigger for the onset of the cosmic acceleration. Such transition occurs at relatively late times ($z_{\rm nr}\sim 4$ for the specific model considered here) 
thereby providing a possible solution to the cosmic coincidence problem. On the other hand, the oscillations of the DE scalar field around the minimum of the potential
determine in turn oscillations of the average neutrino mass according to Eq.~\ref{eq:nu_mass}, that have very significant implications for the evolution of perturbations in the context of Growing
Neutrino cosmologies, and that might provide a direct way to test, constrain, or disproof the model, as we will discuss in full detail in the next Sections.

\section{N-Body simulations}

The linear evolution of perturbations in the context of Growing Neutrino scenarios has been extensively studied by several authors \citep[see e.g.][]{Mota_etal_2008,
Pettorino_etal_2010}, and some attempts have also been made to extend such investigations to the nonlinear regime by means of suitably modified
spherical collapse numerical codes \citep[e.g. by][]{Wintergerst_etal_2010}. A common prediction of all these studies is that the neutrino fluctuations grow very rapidly after the transition redshift
$z_{\rm nr}$ due to the fifth-force mediated by the scalar field which is a factor $2\beta ^{2}$ (i.e. around $5\times 10^{3}$ times for our specific model) 
stronger than the standard gravitational interaction. This rapid growth leads to the formation of very large neutrino structures at scales beyond $\sim 10$ Mpc/$h$
that grow nonlinear at $z\approx 2$. After this time the linear approximation no longer holds, and a full nonlinear treatment of the system is therefore required.
The attempt to use modified spherical collapse algorithms, however, only mildly improves the situation by allowing to follow the evolution of nonlinear bound structures 
up to the onset of virialization, which happens at $z\approx 1.5$ for neutrino halos of size $\sim 10$ Mpc. This further limitation is mainly due to another peculiar effect of 
interacting DE models, i.e. the presence of an extra friction term for the coupled particles arising from momentum conservation. 
After the onset of virialization the system can therefore be studied only by means of suitable N-body simulations, that must include in their implementation all the specific features of
interacting DE models, and in particular of Growing Neutrino scenarios. To this end, we have made use of the specific modification by \citet{Baldi_etal_2010}
of the widely used parallel N-body code {\small GADGET} \citep{gadget-2} to run a series of large N-body simulations for the specific Growing Neutrino model
described above. 
By using such numerical setup we have been able to extend the study of nonlinear
neutrino structures down to $z=1$, and to investigate the growth and the abundance of such structures in a realistic cosmological realization.
The limit of $z=1$ is due to the fact that neutrino velocities grow very rapidly into the relativistic regime
due to their very large accelerations, and the Newtonian approximation on which N-body codes are based becomes not appropriate for such large velocities.

\section{Results}

The results of our numerical simulations have shown for the first time how large neutrino overdensities are seeded by the underlying Cold Dark Matter gravitational potential wells at the
transition redshift $z_{\rm nr}$ and how these bound structures subsequently grow nonlinearly and merge to form isolated neutrino lumps at scales beyond $10$ Mpc.

\subsection{Neutrino large-scale structures}
\begin{figure*}
\includegraphics[scale=0.24]{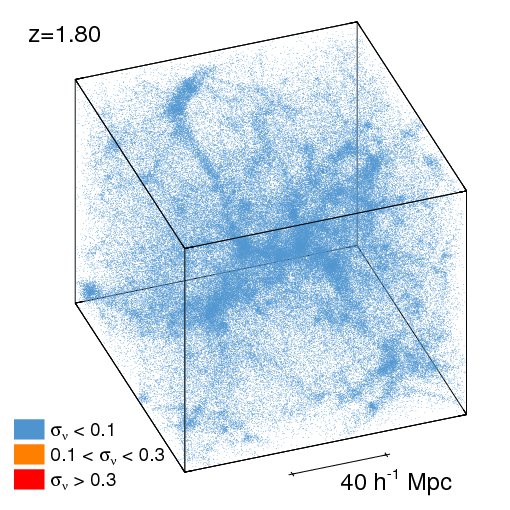}
\includegraphics[scale=0.24]{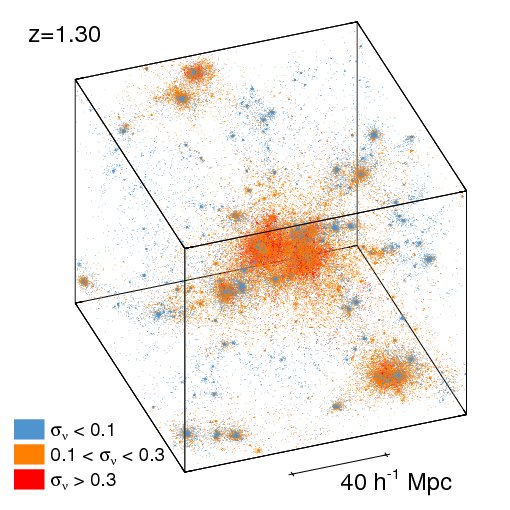}
\includegraphics[scale=0.24]{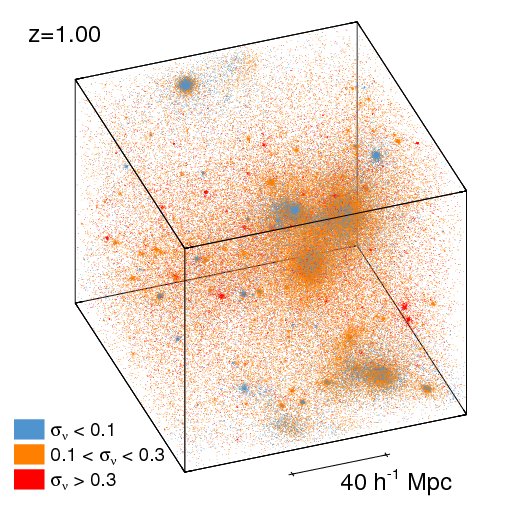}
\caption{Neutrino structures in a box of 120 Mpc/$h$. See \citet{Baldi_etal_2011a} for further details.}
\label{fig:boxes}
\end{figure*}
\normalsize

The evolution of the neutrino distribution at different redshifts in a periodic cosmological box of $120$ Mpc$/h$ aside is shown in Fig.~\ref{fig:boxes}. 
The rapid growth and merging history of neutrino lumps between $z=1.8$ and $z=1.3$ is clearly displayed by comparing the left and the central panels.
However, the right panel shows a significantly lower number of neutrino halos if compared to the situation at $z=1.3$ and correspondingly a large fraction
of free neutrinos that fill the whole simulation volume and do not belong to any bound structure. This peculiar evolution represents one of the main findings of the present work
and can be explained in terms of the rapid oscillations of the neutrino mass due to the oscillations of the DE scalar field $\phi $ around the minimum of its effective potential:
the change in sign of the time derivative of the neutrino mass $\dot{m}_{\nu }$ determined by the oscillations implies a sort of ``pulsation" of the gravitational potential
associated to any given neutrino overdensity. In addition to this, the change in sign of the extra friction term also contributes to trap neutrino particles in bound structures whenever
$\dot{m}_{\nu }>0$ while it favors the escape of neutrino particles from bound objects whenever $\dot{m}_{\nu }<0$ \citep[see][for more details]{Baldi_etal_2011a}. 
We have therefore shown for the first time that nonlinear neutrino lumps forming in the context of Growing Neutrino cosmologies are expected to ``pulsate" following the
oscillations of the DE scalar field.

\subsection{Neutrino halo mass function}
\begin{figure}
\includegraphics[scale=0.3]{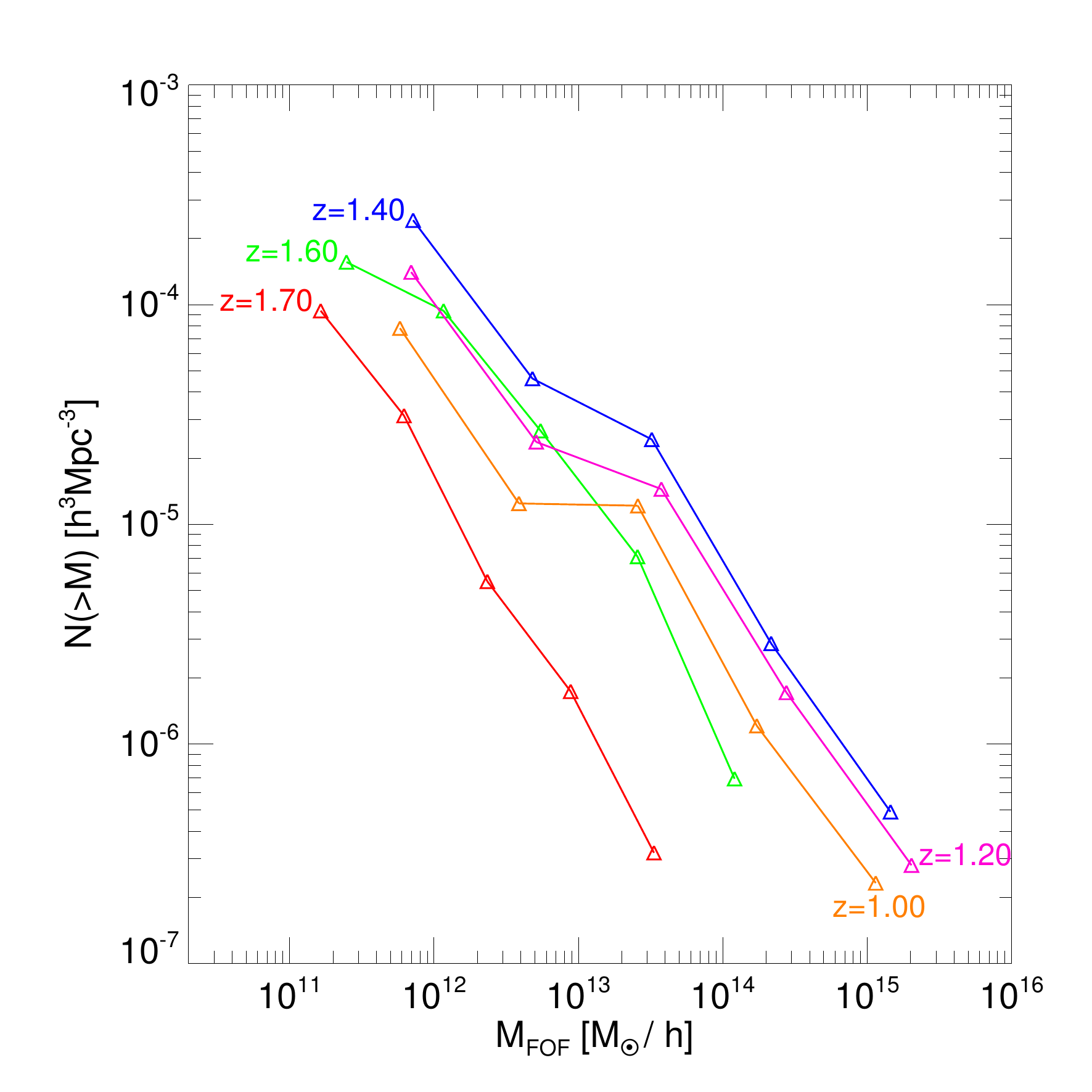}
\caption{The neutrino halo mass function at different redshifts.}
\label{fig:massfunction}
\end{figure}
\normalsize

In order to quantify the expected observational signatures of nonlinear neutrino lumps it is necessary to determine, besides the specific time evolution of any individual
neutrino halo, also a realistic statistical sample of such structures, in order to estimate their volume and line-of-sight densities. This is another issue that
previous investigations could not properly address, and that has found
a first assessment only with the present work.
We have in fact identified neutrino halos in our simulations by means of a suitable modification of the widely used Friends-of-Friends algorithm and we have therefore 
computed a neutrino halo mass function (analogous to the standard CDM halo mass function) over the whole range of masses covered by our numerical simulations. The results of this procedure are shown in Fig.~\ref{fig:massfunction}, where 
the abundance of neutrino halos as a function of mass is plotted for different redshifts. Also this plot clearly shows how the abundance of neutrino halos
first increases in time up to $z\sim 1.4$ and subsequently decays again, due to the scalar field oscillations and the consequent oscillations of the neutrino mass.

\section{Conclusions}

We have presented the results of the first N-body simulations of structure formation for Growing Neutrino cosmologies. Our results have shown for the first time how 
large neutrino lumps forming in the context of such cosmological models show a ``pulsation" due to the rapid oscillations of the DE scalar field around the minimum of its
effective potential. We have also determined, for the first time, a neutrino halo mass function that represents a fundamental quantity in order to put reliable observational
constraints on this class of cosmological models.

\acknowledgments 
I am deeply thankful to Valeria Pettorino, Luca Amendola, and Christof Wetterich, co-authors of the paper on which this presentation is based.

\bibliography{baldi_bibliography} 

\end{document}